\newcommand{\Tr}{\mathop{\mathrm{Tr}}\nolimits} 
\newcommand{\tr}{\mathop{\mathrm{tr}}\nolimits} 
\newcommand{\diag}{\mathop{\mathrm{diag}}\nolimits} 
\newcommand{\rank}{\mathop{\mathrm{rank}}\nolimits} 
\newcommand{\openone}{\leavevmode\hbox{\small1\normalsize\kern-.33em1}} 
\newcommand{\prim}{\sigma}
\newcommand{\basis}{\theta}
\newcommand{\Gal}[1]{\mathrm{GF}(#1)}

\documentclass[seceqn]{elsart1p}
\usepackage{bm,url,cite,amsfonts,amssymb,graphicx,times}

\begin{document}

\begin{frontmatter}
\title{Discrete phase-space structure of $n$-qubit mutually unbiased bases}

\author{A.~B.~Klimov}

\address{Departamento de F\'{\i}sica,
Universidad de Guadalajara, 44420~Guadalajara, Jalisco, Mexico}

\author{J.~L.~Romero}
\address{Departamento de F\'{\i}sica,
Universidad de Guadalajara, 44420~Guadalajara, Jalisco, Mexico}

\author{G.~Bj\"{o}rk}
\address{School of Information and
Communication Technology, Royal Institute of Technology (KTH),
Electrum 229, SE-164 40 Kista, Sweden}

\author{L.~L.~S\'anchez-Soto}
\address{Departamento de \'Optica,
Facultad de F\'{\i}sica, Universidad Complutense,
28040~Madrid, Spain}

\date{\today}

\begin{abstract}
  We work out the phase-space structure for a system of $n$ qubits.
  We replace the field of real numbers that label the axes of the
  continuous phase space by the finite field $\Gal{2^n}$ and
  investigate the geometrical structures compatible with the notion of
  unbiasedness. These consist of bundles of discrete curves
  intersecting only at the origin and satisfying certain additional
  properties. We provide a simple classification of such curves and
  study in detail the four- and eight-dimensional cases, analyzing
  also the effect of local transformations. In this way, we provide a
  comprehensive phase-space approach to the construction of mutually
  unbiased bases for $n$ qubits.
\end{abstract}
\end{frontmatter}

\maketitle

\section{Introduction}

Quantum mechanics describes physical systems through the density
operator $\hat{\varrho}$. For continuous systems, this operator lives
in an infinite-dimensional complex Hilbert space and its relations
with the physical properties of the system are far from obvious. To
overcome these conceptual difficulties, a number of phase-space
methods have been devised, which result in a striking formal
similarity with classical mechanics~\cite{Schleich2001}.

Textbook examples of this subject are usually presented in terms of
continuous variables, typically position and momentum. However, there
are many quantum systems that can be appropriately described in a
finite-dimensional Hilbert space. These include, among other, spins,
multilevel atoms, optical fields with a fixed number of photons,
electrons or molecules with a finite number of sites, etc.  An elegant
way of approaching these systems was proposed by Weyl~\cite{Weyl1950}
in his description of quantum kinematics as an Abelian group of ray
rotations. Similar results were also obtained by
Schwinger~\cite{Schwinger1960a,Schwinger1960b,Schwinger1960c}, who
showed that a set of unitary operators (defined through cyclic
permutations of state vectors) can be constructed such that they are
the generators of a complete operator basis, in terms of which all
possible quantities related to the physical system can be built.

The structure of the phase space associated to a $d$-dimensional
system (a qudit, in the modern parlance of quantum information) 
has been addressed by a number of authors. A possible approach 
was taken by Hannay and Berry~\cite{Hannay1980}, considering a 
grid constrained to admit only periodic probability distributions, 
which implies that it is effectively a $2d \times 2d$-dimensional 
torus.  The same strategy was adopted by 
Leonhardt~\cite{Leonhardt1995,Leonhardt1996} and used to deal 
with different aspects of quantum
information~\cite{Miquel2002a,Miquel2002b,Paz2002}.  This method 
offers a way for treating even-dimensional systems, since the 
grid has both integer and half-odd coordinates.

However, the mainstream of research has focused on a phase space
pictured as a $d \times d$ lattice.  This line was started by
Buot~\cite{Buot1973}, who introduced a discrete Weyl transform that
generates a Wigner function on the toroidal lattice $\mathbb{Z}_d$
(with $d$ odd). More recently, these ideas have been developed further
by other authors~\cite{Wootters1987,Galetti1988,Cohendet1988,
Galetti1992,Kasperkovitz1994,Opatrny1995,Galetti1996,Opatrny1996,
Rivas1999,Gibbons2004,Vourdas2005,Vourdas2007}.  In particular, 
when the dimension is a power of a prime, one can label the points 
in the $d\times d$ grid with elements of the finite Galois field
$\Gal{d}$. At first sight, the use of elements of $\Gal{d}$ as
coordinates could be seen as an unnecessary complication, but it 
turns out to be an essential step: only by doing this we can endow 
the phase space with similar geometrical properties as the ordinary 
plane.  Note also that though the restriction to powers of primes 
rules out many quantum systems, this formulation is ideally suited 
for the outstanding case of $n$ qubits we deal in this paper.

In these finite descriptions, the Wigner function, being the Weyl
representative of the density operator, naturally emerges as a
function that takes values only at the points defining the discrete
mesh of the phase space (while preserving some properties that make it
a special object in quantum mechanics). A remarkable feature is that
one can sum the Wigner function along different axes (including skew
ones) to obtain correct probability distributions for observables
associated with those axes. Although the axis observables cannot
be complementary in the usual sense (their commutator cannot be
proportional to the identity operator), they will have a closely
related property: every eigenstate of either one of them is a state of
maximum uncertainty with respect to the other.

This makes a deep connection with the notion of mutually unbiased
bases (MUBs)~\cite{Delsarte1975,Wootters1986}, which were introduced
as a central tool for quantum state reconstruction~\cite{Wootters1989}. 
They also play a relevant role in a proper understanding of
complementarity~\cite{Kraus1987,Lawrence2002,Chaturvedi2002,Petz2007}, in
cryptographic protocols~\cite{Bechmann-Pasquinucci2000,Asplund2001},
and in quantum error correction codes~\cite{Gottesman1996,Calderbank1997}.
Recently, they have also found uses in quantum game theory, in particular 
to provide a convenient tool for solving the so-called mean king 
problem~\cite{Aharonov2001,Englert2001,Aravind2003,Hayashi2005,
Klappenecker2005,Paz2005,Durt2006,Kimura2006}.

It has been shown~\cite{Ivanovic1981} that the maximum number of 
MUBs can be at most $d+1$. Actually, it is known that if $d$ is 
prime or power of a prime the maximal number of MUBs can be
achieved~\cite{Calderbank1997a}.  Different explicit constructions 
of MUBs in prime power dimensions have been suggested in a number of
recent papers~\cite{Klappenecker2003,Lawrence2004,Parthasarathy2004,
Pittenger2005,Durt2005,Planat2005a,Klimov2005,Boykin2007}. Remarkably
though, there is no known answer for any other values of $d$, although
there are some attempts to find a solution in some simple 
cases, such as $d=6$ or when $d$ is a nonprime integer
squared~\cite{Grassl2004,Archer2005,Wocjan2005,Butterley2007}.
Recent work has suggested that the answer to this question may 
well be related with the nonexistence of finite projective 
planes of certain orders~\cite{Saniga2004,Bengtsson2004} or with 
the problem of mutually orthogonal Latin squares in
combinatorics~\cite{Zauner1999,Wootters2006}.

The construction of MUBs is closely related to the possibility of
finding $d+1$ disjoint classes, each one having $d-1$ commuting
operators [which proves useful to arrange in a table with $(d-1)
\times (d+1)$ entries], so the corresponding eigenstates form sets 
of MUBs~\cite{Bandyopadhyay2002}. Nevertheless, these MUB operators 
can be organized in several different nontrivial tables, leading to
different factorization properties of the MUB~\cite{Romero2005}.  
It has been recently noticed~\cite{Bjork2007} that such arrangements 
are related with special types of curves.  We have previously
analyzed~\cite{Klimov2007} those curves in the four-dimensional case
(corresponding to two qubits) and have shown that they can be obtained
through local transformations from rays (straight lines passing though
the origin), so that the six possible $3 \times 5$ tables of operators
lead to the same (and unique) factorization of the MUBs.

In the present paper we go further and analyze the general
situation of $n$ qubits. In particular, we classify the admissible
curves in specific bundles and show how the properties of these curves 
can be used to determine nontrivial sets of MUBs.

\section{Mutually unbiased bases and discrete phase space}
\label{MUBs}

\subsection{Constructing mutually unbiased bases in prime dimensions}

We start by considering a system living in a Hilbert space
$\mathcal{H}_{d}$, whose dimension $d$ is assumed for now to be a
prime number $p$. The different outcomes of a maximal test constitute
an orthogonal basis of $\mathcal{H}_{d}$~\cite{Peres1993}.  One can also 
look for orthogonal bases that, in addition, are ``as different as possible''. 
This is the idea behind MUBs and can be formally stated as follows: two
bases $\{ | u_i \rangle \}$ and $\{ | v_j \rangle \}$ are mutually
unbiased when
\begin{equation}
  \label{eq:defMUBS}
| \langle u_i | v_j \rangle|^2 = \frac{1}{d} \, . 
\end{equation}
Unbiasedness also applies to measurements: two nondegenerate tests 
are mutually unbiased if the bases formed by their eigenstates are MUBs.
Therefore, the measurements of the components of a spin 1/2 along $x$, $y$, 
and $z$ axes are all unbiased. It is also obvious that for these finite 
quantum systems unbiasedness is tantamount of complementarity.

It is useful to choose a computational basis $| n \rangle $ ($n = 0, \ldots , 
d-1$) in $\mathcal{H}_{d}$ and introduce the basic operators
\begin{equation}
\label{CC}
X | n \rangle =  |n + 1 \rangle \, ,
\qquad \qquad
Z | n \rangle  =  \omega^{n} | n \rangle ,
\end{equation}
where $ \omega = \exp (2\pi i/d) $ is a $d$th root of the unity 
and addition and multiplication must be understood modulo $d$.  
These operators $X$ and $Z$, which are generalizations of the 
Pauli matrices, were studied long ago by Weyl~\cite{Weyl1950} 
and have  been used recently by many authors in a variety of
applications~\cite{Gottesman2001,Bartlett2002}.  They generate 
a group under multiplication known as the generalized Pauli group 
and obey
\begin{equation}
  Z X = \omega \,  X Z ,
\end{equation}
which is the finite-dimensional version of the Weyl form of the
commutation relations.

As anticipated in the Introduction, one can construct MUBs by finding
$d+1$ disjoint classes (each one having $d-1$ commuting operators),
such that the corresponding eigenstates form sets of MUBs. We follow
the explicit construction in reference \cite{Klimov2005}, which starts
with the following sets of operators:
\begin{equation}
\label{SCop}
\{ X^{k} \} ,
\quad 
\{ Z^{k} X^{mk} \} ,
\end{equation}
with $k = 1, \ldots , d-1$ and $m = 0, \ldots , d-1$.
One can easily check that
\begin{eqnarray}
\label{pairwise}
\Tr ( X^k X^{k^\prime}{}^\dagger ) =
d \, \delta_{k, k^\prime} ,
\qquad
\Tr ( Z^k Z^{k^\prime} {}^\dagger ) =
d \, \delta_{k, k^\prime} ,  \nonumber  \\
 \\
\Tr [ (Z^{k} X^{mk})  ( Z^{k^\prime} X^{m^\prime k^\prime})^\dagger ] = 
d \, \delta_{k, k^\prime}
\delta_{m, m^\prime} .
\nonumber
\end{eqnarray}
These pairwise orthogonality relations indicate that, for every value
of $m$, we generate a maximal set of $d-1$ commuting operators and
that all these classes are disjoint.  In addition, the common
eigenstates of each class $m$ form different sets of MUBs.

\subsection{Mutually unbiased bases for $n$ qubits}

When the space dimension $d=p^{n}$ is a power of a prime it is natural
to view the system as composed of $n$ constituents (particles), each
of dimension $p$.  We adapt the previous construction to this case,
although with an eye on the particular case of $n$ qubits (in the Appendix
we summarize the basic notions of finite fields needed for our purposes
in this paper).

The main idea consists in labeling the states with elements of the
finite field $\Gal{2^n}$, instead of natural numbers.  We denote by
$|\alpha \rangle$, with $\alpha \in \Gal{2^n}$, an orthonormal basis
in the Hilbert space of the system.  Operationally, the elements of
the basis can be labeled by powers of a primitive element, and the
basis reads
\begin{equation}
  \{|0 \rangle, \, |\prim \rangle, \ldots, \,|\prim^{2^n-1} = 1 \rangle \} .
\end{equation}
These vectors are eigenvectors of the operators $Z_{\beta}$ belonging 
to the generalized Pauli group, whose generators are now defined as
\begin{equation}
  Z_{\beta}  =  \sum_{\alpha \in \Gal{2^n}} \chi ( \alpha \beta ) \,
  | \alpha \rangle \langle \alpha |,
  \qquad \qquad
  X_{\beta} = \sum_{\alpha \in \Gal{2^n}}
  | \alpha + \beta \rangle \langle \alpha |, \label{XZgf} \\
\end{equation}
so that
\begin{equation}
  Z_{\alpha} X_{\beta} = \chi ( \alpha \beta ) \, X_{\beta} Z_{\alpha},
\end{equation}
where $\chi$ is an additive character defined in (\ref{Eq: addchardef}).

The operators (\ref{XZgf}) can be factorized into tensor products of powers
of single-particle Pauli operators $\sigma_{z}$ and $\sigma_{x}$, whose
expression in the standard basis of the two-dimensional Hilbert space is
\begin{equation}
  \sigma_{z} = | 1 \rangle \langle 1 | - |0 \rangle \langle 0 |,
  \qquad \qquad
  \sigma_{x} = | 0 \rangle \langle 1 | + | 1 \rangle \langle 0 | .
  \label{sigmas}
\end{equation}
This factorization can be carried out by mapping each element of
$\Gal{2^n}$ onto an ordered set of natural numbers as in
equation~(\ref{mapnum}).  A convenient choice for this is the selfdual
basis, since the finite Fourier transform factorizes then into a
product of single-particle Fourier operators, which leads to
\begin{equation}
  Z_{\alpha}  = \sigma_{z}^{a_{1}} \otimes \ldots \otimes
  \sigma_{z}^{a_{n}},
  \qquad \qquad
  X_{\beta} = \sigma_{x}^{b_{1}} \otimes \ldots \otimes
  \sigma_{x}^{b_{n}} 
\end{equation}
where $(a_{1}, \ldots, a_{n})$ and $(b_{1}, \ldots, b_{n})$ are
the corresponding coefficients.

The simplest geometrical structures in the discrete phase space are
straight lines, i.e., collections of points $( \alpha , \beta )
\in \Gal{2^n} \times \Gal{2^n}$ satisfying the relation
\begin{equation}
  \zeta \alpha + \eta \beta = \vartheta ,
  \label{line}
\end{equation}
where $\zeta ,\eta$ and $\vartheta $ are fixed elements of
$\Gal{2^n}$.  Two lines
\begin{equation}
  \zeta \alpha + \eta \beta = \vartheta ,
  \qquad \qquad
  \zeta^{\prime} \alpha + \eta^{\prime} \beta = \vartheta^{\prime},
  \label{line1}
\end{equation}
are parallel if they have no common points, which implies that $\eta
\zeta^{\prime} = \zeta \eta^{\prime}$. If the lines (\ref{line1}) are
not parallel they cross each other. A ray is a line passing through
the origin, so its equation is
\begin{equation}
  \alpha = 0 ,
  \qquad
  \mathrm{or}
  \qquad
  \beta = \lambda \alpha ,
  \label{rays}
\end{equation}
or, in parametric form,
\begin{equation}
  \alpha (\kappa ) = \mu \kappa ,
  \qquad \qquad
  \beta (\kappa )= \nu \kappa ,
  \label{ray_pf}
\end{equation}
where $\kappa$ is the parameter running through the field. The rays
are the simplest nonsingular (i.e., with no selfintersection) additive
structures in phase space, in the sense that
\begin{eqnarray}
  \label{ac}
  \alpha (\kappa + \kappa^{\prime}) = \alpha ( \kappa ) +
  \alpha (\kappa^{\prime}),
  \qquad \qquad
  \beta (\kappa + \kappa^{\prime}) = \beta (\kappa ) +
  \beta (\kappa^{\prime}) .
\end{eqnarray}
This means that by summing the coordinates of the origin and of any
point in a ray we obtain another point on the same ray. In particular,
this opens the possibility of introducing operators that generate
``translations'' along these rays~\cite{Gibbons2004}.

The rays have a very remarkable property: the monomials
$Z_{\alpha}X_{\beta}$ (labeled by phase-space points) belonging 
to the same ray commute
\begin{equation}
  Z_{\alpha_{1}} X_{\beta_{1} = \lambda \alpha_{1}} \,
  Z_{\alpha_{2}} X_{\beta_{2} = \lambda \alpha_{2}}
  =
  Z_{\alpha_{2}} X_{\beta_{2} = \lambda \alpha_{2}} \,
  Z_{\alpha_{1}} X_{\beta_{1} = \lambda \alpha_{1}},
\end{equation}
and thus, have a common system of eigenvectors 
$\{|\psi_{\upsilon,\lambda} \rangle \}$, with $\lambda ,\upsilon \in
\Gal{2^n}$:
\begin{equation}
  Z_{\alpha} X_{\lambda \alpha} | \psi_{\upsilon,\lambda} \rangle =
  \exp(i \xi_{\upsilon, \lambda}) | \psi_{\upsilon, \lambda} \rangle ,
  \label{ZXes}
\end{equation}
where $\lambda $ is fixed and $\exp (i \xi_{\upsilon, \lambda})$ is
the corresponding eigenvalue, so  $| \psi_{\upsilon, 0} \rangle =
| \upsilon \rangle $ are eigenstates of $Z_{\alpha}$ (displacement
operators labeled by the ray $\beta =0$, which we take as the 
horizontal axis). Indeed, we have that
\begin{equation}
  \label{eq:unbias}
  |\langle \psi_{\upsilon, \lambda} | 
  \psi_{\upsilon^\prime, \lambda^\prime} \rangle |^2 =
  \delta_{\lambda, \lambda^\prime} \delta_{\upsilon, \upsilon^\prime}
  + \frac{1}{d} (1 - \delta_{\lambda, \lambda^\prime}),
\end{equation}
and, in consequence, they are MUBs~\cite{Wootters1989}.  Since 
each ray defines a set of $2^{n}-1$ commuting operators, if we 
introduce $2^{n}+1$ sets of commuting operators (which from now on
will be called displacement operators) as
\begin{equation}
  \{X_{\beta}\}, \quad \{Z_{\alpha} X_{\beta =\lambda \alpha}\} \, ,
\label{set1}
\end{equation}
then we have a whole bundle of $2^{n}+1$ rays (which is obtained
by varying the ``slope'' $\lambda$) that allows us to construct a
complete set of MUB operators arranged in a $(2^{n}-1) \times
(2^{n}+1)$ table.

We wish to emphasize that in our approach we do not assign a 
quantum state to each line in phase space (as in
reference~\cite{Gibbons2004}), but rather we use them to label 
Pauli displacement operators.

\section{Curves in discrete phase space}
\label{Sec: Curves}

\subsection{Additive curves and the commutativity condition}

The rays are not the only additive structures that exist in the
discrete phase space. One can check that the parametric curves 
(passing through the origin)
\begin{equation}
  \alpha (\kappa ) = \sum_{m=0}^{n-1} \alpha_{m} \, \kappa^{2^m},
  \qquad \qquad
  \beta (\kappa ) = \sum_{m=0}^{n-1} \beta_{m} \, \kappa^{2^m},
  \label{curve1}
\end{equation}
satisfy the condition (\ref{ac}) too. If we also require the 
displacement operators labeled with the points of (\ref{curve1}) 
to commute with each other, we must impose
\begin{equation}
  \tr (\alpha \beta^{\prime} ) = \tr (\alpha^{\prime} \beta ),
  \label{Tr_cond}
\end{equation}
where $\alpha^{\prime} = \alpha (\kappa^{\prime})$ and $\beta^{\prime}
= \beta ( \kappa^{\prime})$.  Then, the coefficients $\alpha_{m}$ and
$\beta_{m}$ fulfill the following restrictions (the indices must be
understood modulus $n$)
\begin{equation}
  \sum_{m=0}^{n-1} \alpha_{n-m}^{2^{m}} \, \beta_{n-m+q}^{2^{m}} =
  \sum_{m=0}^{n-1}\beta_{n-m}^{2^{m}} \, \alpha_{n-m+q}^{2^{m}},
  \qquad q=1, \ldots ,n-1 .
  \label{cc1}
\end{equation}
This can be rewritten in an invariant form by summing up all 
Frobenius automorphisms of (\ref{cc1}):
\begin{equation}
  \sum_{m \neq k} \tr (\alpha_{m} \beta_{k}) = 0 .
  \label{cc_g2}
\end{equation}
Whenever the condition (\ref{cc_g2}) holds true, we can associate to
each curve (\ref{curve1}), with given coefficients $\vec{\alpha} =
(\alpha_0, \ldots, \alpha_{n-1})$ and $\vec{\beta} = (\beta_0, \ldots,
\beta_{n-1})$, a state $| \psi_{\vec{\alpha}, \vec{\beta}} \rangle$.

The curves fulfilling equation~(\ref{cc_g2}) will be called additive
commutative curves. When such a curve contains $2^{n}-1$ different 
points (apart from the origin), the monomials $Z_{\alpha} X_{\beta}$ 
form also a set of commuting operators (as happened for the rays). 

In fact, consider a subset  $Z_{\alpha (\kappa )} X_{\beta (\kappa )}$, 
such that 
\begin{equation}
  \label{eq:cond}
  [ Z_{\alpha (\kappa)} X_{\beta (\kappa)}, 
  Z_{\alpha (\kappa^{\prime})} X_{\beta (\kappa^{\prime})} ] = 0,
  \qquad \qquad
  \Tr ( Z_{\alpha (\kappa)} X_{\beta (\kappa)} 
  Z_{\alpha (\kappa^{\prime})}^\dagger X_{\beta(\kappa^{\prime})}^\dagger ) = 0, 
\end{equation}
for any $\kappa ,\kappa^\prime \in \Gal{p^{n}}$, i.e. a disjoint set of 
$d$ mutually commuting monomials, including $Z_{\alpha(0)} 
X_{\beta(0)} = \openone$. Then, the eigenstates of any two disjoint sets 
of these mutually commuting monomials form MUBs.

To prove this important result, we first note that any set of commuting 
monomials can be obtained by applying Clifford operations $U_{\lambda}$ 
to the simplest set $\{ Z_{\kappa} \}$: 
\begin{equation}
  U_{\lambda} Z_{\kappa} U_{\lambda}^{\dagger} = 
  \phi (\lambda ,\kappa) \, D_{\lambda, \kappa} ,  
  \label{U_d}
\end{equation}
where $\phi (\lambda , \kappa )$ is an unessential phase factor 
[$\phi(\lambda ,0) = \phi (0, \kappa ) = 1$] and  
$D_{\lambda, \kappa} = Z_{\alpha(\lambda ,\kappa )} X_{\beta (\lambda ,\kappa )}$ 
fulfill
\begin{equation} 
  [ D_{\lambda, \kappa}, D_{\lambda, \kappa^{\prime}} ] = 0 , 
  \qquad \qquad 
  \Tr ( D_{\lambda, \kappa} D_{\lambda,\kappa^{\prime}}^{\dagger} ) = 
  d \, \delta_{\kappa, \kappa^{\prime}} ,
\end{equation}
since the transformations $U_{\lambda }$ are nondegenerate.
For two disjoint set of commuting monomials 
$\{ D_{\lambda, \kappa} \}$ and $\{D_{\lambda^\prime, \kappa^\prime} \}$
we have then 
\begin{equation}
  \Tr  ( D_{\lambda, \kappa} D_{\lambda^\prime, \kappa^{\prime}}^\dagger ) =
  \left \{ 
    \begin{array}{cc}
      d \, \delta_{\kappa, \kappa^{\prime}},  & \quad
      \lambda =\lambda^{\prime} , \\ 
      d \, \delta_{\kappa, 0} \delta_{\kappa^{\prime}, 0}, & \quad
      \lambda \neq \lambda^{\prime}
    \end{array}
  \right . .   
  \label{disj_cond2}
\end{equation}
Let $\{ | \upsilon \rangle \}$ be the basis of eigenstates of $Z_{\kappa }$. 
It is worth noting the following expansion
\begin{equation}
  \label{al}
  |\upsilon \rangle \langle \upsilon | = 
  \frac{1}{d}  \sum_{\kappa} \chi (- \upsilon \kappa ) Z_{\kappa} = 
  \frac{1}{d}  \sum_{\kappa} \chi (\upsilon \kappa ) Z_{\kappa}^{\dagger}.
\end{equation}
Now, if we define the states $ |\psi_{\upsilon, \lambda} \rangle = U_{\lambda} 
| \upsilon \rangle$ and $|\psi_{\upsilon^{\prime},\lambda^{\prime}} \rangle = 
U_{\lambda^{\prime}}  |\upsilon^{\prime} \rangle$, where $U_{\lambda}$ is as in 
equation~(\ref{U_d}), then a direct calculation using (\ref{al}) shows
that (\ref{eq:unbias}) holds true for them and so they are indeed MUBs,
as announced. 

Finding sets of MUBs can be thus reduced to the problem of arranging
additive curves in $2^{n}+1$ bundles of mutually nonintersecting curves.

Due to the condition (\ref{ac}), points of a curve form a
finitely generated Abelian group, which allows us to determine 
all the curve from any $n$ points and, in particular, from the ``first'' 
$n$ consecutive points. For instance, taking the parameter $\kappa $
polynomially ordered (that is, $\kappa = \prim,\prim^{2}, \ldots,
\prim^{2^{n}-1}$), we have that $\alpha (\prim^{k}) + \alpha
(\prim^{k+1}) = \alpha (\prim^{k} + \prim^{k+1}) = \alpha [\prim^{k}
(1 + \prim)] = \alpha [\prim^{k+ L(1)}],$ where $L(\lambda )$ is the
Jacobi logarithm~\cite{Lidl1986}. Given a primitive polynomial,
sometimes is possible to evaluate $L(1)$ in a simple form.

For $\Gal{2^2}$ the only irreducible polynomial is $x^{2} + x + 1 =
0$, and this immediately leads to $L(1)= L(\prim^{3}) = 2$, in such a
way that $\alpha (\prim^{k}) + \alpha (\prim^{k+1}) =\alpha
(\prim^{k+2})$ and two points are enough to determine any additive
curve. In the case of $\Gal{2^3}$, if we use the irreducible
polynomial $x^{3} + x + 1 = 0$, we obtain that $L(1) =
L(\prim^{7})=3$, so that $\alpha (\prim^{k}) + \alpha (\prim^{k+1}) =
\alpha (\prim^{k+3})$ and we need three points to generate the curve.

\subsection{Nonsingularity condition}
\label{Sec: Nonsingularity}

To classify all the additive commutative curves, we first discuss the
condition of nonsingularity (i.e., nonselfintersection), which means
that there are no $\kappa ^{\prime }\neq \kappa $ such that
\begin{equation}
  \alpha (\kappa ) = \alpha ( \kappa^{\prime}),
  \qquad \qquad
  \beta (\kappa ) = \beta (\kappa^{\prime}).
  \label{sc0}
\end{equation}
If one of these equations would be fulfilled, then $\mathfrak{S}^{m} [
\varepsilon (\kappa ) ]=0$ (with $\varepsilon = \alpha$ or $\beta$)
for $m=0, \ldots , n-1$.

Let us introduce the following matrix that will play a relevant
role in our subsequent analysis:
\begin{equation}
  \label{W}
  \mathsf{W}_{\vec{\varepsilon}} = 
  \left (
    \begin{array}{cccc}
      \varepsilon_{0} & \varepsilon_{1} & \varepsilon_{2} & \ldots  \\
      \varepsilon_{n-1}^{2} & \varepsilon_{0}^{2} & \varepsilon_{1}^{2} & \ldots \\
      \varepsilon_{n-2}^{4} & \varepsilon_{n-1}^{4} & \varepsilon_{0}^{4} & \ldots  \\
      \vdots & \vdots & \vdots & \ddots
    \end{array}
  \right ) ,
  \label{MN}
\end{equation}
where the rows are determined by the coefficients in (\ref{curve1})
and the corresponding expansions of $\mathfrak{S}^{m} [ \varepsilon (\kappa)]$ 
(with $\varepsilon = \alpha, \beta$ in our case). If $\det 
\mathsf{W}_{\vec{\alpha}}$ and/or $\det \mathsf{W}_{\vec{\beta}}$ 
do not vanish simultaneously, the curve (\ref{curve1}) has no 
selfintersection.

The condition
\begin{equation}
  \det \mathsf{W}_{\vec{\alpha}} = \det \mathsf{W}_{\vec{\beta}} = 0 ,
  \label{sc1}
\end{equation}
indicates that the ranks of the matrices $\mathsf{W}_{\vec{\alpha}}$
and $\mathsf{W}_{\vec{\beta}}$ are smaller than the dimension of the
system, but it does not guarantee that there exist $\kappa^{\prime}
\neq \kappa$ satisfying (\ref{sc0}), because the solutions of $\alpha
(\kappa ) = \alpha (\kappa^{\prime})$ and $\beta (\kappa^{\prime
  \prime}) = \beta (\kappa^{\prime \prime \prime})$ can form disjoint
sets.  Consequently, (\ref{sc1}) is necessary but not sufficient to
determine if a curve is singular. Another necessary but not sufficient
condition of singularity is $\det (\mathsf{W}_{\vec{\alpha}} +
\mathsf{W}_{\vec{\beta}} )= 0$.

A curve that fulfills $\det \mathsf{W}_{\vec{\alpha}} \neq 0$ and/or
$\det \mathsf{W}_{\vec{\beta}} \neq 0$ will be called a regular
curve. For such curves the coordinate $\alpha $ (when $\det
\mathsf{W}_{\vec{\alpha}} \neq 0$) or $\beta $ (when $\det
\mathsf{W}_{\vec{\beta}} \neq 0$) take all the values in the field.

Nonsingular curves satisfying (\ref{sc1}) will be called exceptional
curves.  The conditions (\ref{sc1}) mean that $\mathfrak{S}^{m}
(\alpha )$ and $\mathfrak{S}^{m} ( \beta )$ are not linearly
independent (for $m=0, \ldots ,n-1$), so neither $\alpha $ or
$\beta $ run through the whole field (in other words, the values of
$\alpha $ and $\beta $ are degenerate). The number of linearly
independent powers of $\alpha $ (respectively $\beta $) is equal to the 
rank of the matrix $\mathsf{W}_{\vec{\alpha}}$ (respectively
$\mathsf{W}_{\vec{\beta}}$) and the quantities $n - \rank
\mathsf{W}_{\vec{\alpha}}$ (respectively $n -\rank \mathsf{W}_{\vec{\beta}}$)
determine the degree of degeneration of every allowed value of $\alpha$ 
(respectively $\beta$).

It is interesting to note that the determinant of the matrix
(\ref{MN}) takes only the values zero and one; i.e., $\det
\mathsf{W}_{\vec{\varepsilon}} \in \mathbb{Z}_{2}$, which can be
easily seen by observing that $ ( \det \mathsf{W}_{\vec{\varepsilon}})^{2} 
= \det \mathsf{W}_{\vec{\varepsilon}}$.

\section{Regular curves}
\label{Sec: Regular}

\subsection{Explicit forms}

Given a regular curve, we can invert one of the relations
(\ref{curve1}) and, by substituting into the other one, we find an
explicit equation of the curve. When $\det \mathsf{W}_{\vec{\alpha}}
\neq 0$ and $\det \mathsf{W}_{\vec{\beta}}\neq 0$, the coordinates 
$\alpha$ and $\beta$ are nondegenerate and the curve can be written 
either as
\begin{equation}
  \beta = \sum_{m=0}^{n-1} \phi_{m} \, \alpha^{2^{m}} ;  \label{b}
\end{equation}
or
\begin{equation}
  \alpha = \sum_{m=0}^{n-1} \psi_{m} \, \beta^{2^{m}}.  \label{a}
\end{equation}
However, when $\det \mathsf{W}_{\vec{\alpha}} \neq 0$ but $\det
\mathsf{W}_{\vec{\beta}}=0$, the coordinate $\beta$ is degenerate and
the curve cannot be expressed in the form~(\ref{a}).  We will refer 
to the corresponding curve as $\alpha$-curve.  Similarly, if $\det
\mathsf{W}_{\vec{\beta}}\neq 0$ but $\det \mathsf{W}_{\vec{\alpha}}=0$, 
the coordinate $\alpha$ is degenerate and the curve cannot be expressed 
in the form~(\ref{b}): we will call it a $\beta$-curve.

\subsection{Commutativity conditions}

The commutativity condition (\ref{cc_g2}) can be further simplified for
regular curves. When $\det \mathsf{W}_{\vec{\alpha}} \neq 0$ (or $\det
\mathsf{W}_{\vec{\beta}} \neq 0)$ we obtain, by direct substitution of
the explicit forms (\ref{b}) or (\ref{a}) into (\ref{Tr_cond}), the
following restrictions on the coefficients $\phi_{m}$ (or $\psi_{m})$
\begin{equation}
   \label{cc}
  \phi_{j} = \phi_{n-j}^{2^{j}},
  \qquad \qquad
  \psi_{j} = \psi_{n-j}^{2^{j}},
  \qquad j = 1, \ldots , [(n-1)/2 ],
\end{equation}
where $[ \ ]$ denotes the integer part. For even values of $n$, 
the additional requirements $\phi_{n/2} = \phi_{n/2}^{2^{n/2}}$
($\psi_{n/2} = \psi_{n/2}^{2^{n/2}}$) should be fulfilled, which
basically implies that $\phi_{n/2}$ ($\psi_{n/2}$) belong to the
subfield $\Gal{2^{n/2}}$.

Because the regular curves are nonsingular per definition, we do not
have to carry out the whole analysis involving the parametric forms of
curves and the properties of the corresponding
$\mathsf{W}_{\vec{\alpha}}$ and $\mathsf{W}_{\vec{\beta}}$, but just
to write down explicit expressions using directly (\ref{cc}).

Two regular curves defined explicitly as $\beta =
f(\alpha,\vec{\phi})$ and $\beta = f (\alpha, \vec{\phi}^{\prime})$,
with $\vec{\phi} = (\phi_{0}, \ldots , \phi_{n-1})$ [or $\alpha =g
(\beta , \vec{\psi})$ and $\alpha =g (\beta , \vec{\psi}^{\prime})$,
with $\vec{\psi} = (\psi_{0}, \ldots , \psi_{n-1})$] are not mutually
intersecting (except at the origin), if
\begin{equation}
  \det ( \mathsf{W}_{\vec{\phi}} + \mathsf{W}_{\vec{\phi}^{\prime}} ) \neq 0,
\qquad \qquad
  \det ( \mathsf{W}_{\vec{\psi}} + \mathsf{W}_{\vec{\psi}^{\prime}} ) \neq 0 , 
\label{nir}
\end{equation}
for $\alpha$- and $\beta$-curves, respectively, and the matrices
$\mathsf{W}$ have been defined in (\ref{W}). 

An $\alpha$-curve $\beta = f (\alpha ,\vec{\phi})$ intersects with a
$\beta$-curve $\alpha = g (\beta , \vec{\psi})$ when the polynomial
$\beta =f [g (\beta , \vec{\psi}), \vec{\phi}]$ (or $\alpha = g [f
(\alpha , \vec{\phi}), \vec{\psi}]$) has at least one nonzero root.

It follows from (\ref{nir}) that the regular curves
\begin{equation}
  \beta = \phi_{0} \alpha + \sum_{m=1}^{n-1} \phi_{m} \alpha^{2^{m}},
\label{b1}
\end{equation}
where $\phi_{m}$ ($m=1, \ldots, n-1$) are fixed and fulfill (\ref{cc}), 
and $\phi_{0}$ runs through the whole field, belong to a bundle of 
nonintersecting curves, since the matrices $\mathsf{W}_{\vec{\phi}} + 
\mathsf{W}_{\vec{\phi}^{\prime}}$  take now the diagonal form
\begin{equation}
  \mathsf{W}_{\vec{\phi}} + \mathsf{W}_{\vec{\phi}^{\prime}} = 
  \diag \left [  \phi_{0} +  \phi_{0}^{\prime},
  ( \phi_{0} + \phi_{0}^{\prime} )^{2} , \ldots ,
  ( \phi_{0} + \phi_{0}^{\prime} )^{2^{n-1}} \right ] ,
\end{equation}
so that
\begin{equation}
  \det ( \mathsf{W}_{\vec{\phi}} + \mathsf{W}_{\vec{\phi}^{\prime}}) =
  ( \phi_{0}+\phi_{0}^{\prime})^{2^{n}-1},
\end{equation}
and thus, $\det ( \mathsf{W}_{\vec{\phi}} + \mathsf{W}_{\vec{\phi}^{\prime}}) 
= 1$  if $\phi_{0} \neq \phi_{0}^{\prime}$. To complete the bundle to $n+1$ 
curves we add the ray $\alpha =0$, which obviously has no common points 
with  (\ref{b1}).

Similarly, the curves
\begin{equation}
  \alpha = \psi_{0} \beta + \sum_{m=1}^{n-1} \psi_{m} \, \beta^{2^{m}}
  \label{a1}
\end{equation}
form bundles of nonintersecting curves, except that now we have to
add the ray $\beta=0$ to complete the bundle.

\subsection{Examples}

\subsubsection{Regular curve}

Let us consider the following parametric curve in $\Gal{2^3}$
\begin{eqnarray}
  \alpha = \prim^{2} \kappa + \kappa^{2} + \prim^{4} \kappa^{4} ,
  \qquad \qquad
  \beta = \prim^{3} \kappa + \prim^{6} \kappa^{2} + \prim^{6} \kappa^{4},
\end{eqnarray}
where $\prim$ is the primitive element. The associated matrices are
\begin{equation}
  \mathsf{W}_{\vec{\alpha}} = \left (
    \begin{array}{ccc}
      \prim^{2} & 1 & \prim^{4} \\
      \prim & \prim^{4} & 1 \\
      1 & \prim^{2} & \prim
    \end{array}
  \right ) ,
  \qquad \qquad
  \mathsf{W}_{\vec{\beta}} = \left (
    \begin{array}{ccc}
      \prim^{3} & \prim^{6} & \prim^{6} \\
      \prim^{5} & \prim^{6} & \prim^{5} \\
      \prim^{3} & \prim^{3} & \prim^{5}
    \end{array}
  \right ) .
\end{equation}
One can check that $\det \mathsf{W}_{\vec{\alpha}}=\det
\mathsf{W}_{\vec{\beta}}=1$, and the explicit forms of the curve are
\begin{equation}
  \beta = \prim^{6} \alpha + \prim^{3} \alpha^{2}
  + \prim^{5} \alpha^{4},
  \quad
  \mathrm{or}
  \quad
  \alpha =\prim^{6} \beta + \prim^{3} \beta^{2}
  + \prim^{5}\beta^{4} ,
\end{equation}
whose coefficients satisfy the condition (\ref{cc}). The set
of commuting operators corresponding to this curve is
\begin{equation}
\{ Z_{\prim^{6}} X_{\prim^{5}},\, Z_{\prim^{5}} X_{\prim^{6}}, \,
Z_{\prim^{4}} X_{\prim^{2}}, \, Z_{\prim} X_{\prim}, \,
Z_{\prim^{7}} X_{\prim^{7}}, Z_{\prim^{2}} X_{\prim^{4}}, \,
Z_{\prim^{3}}X_{\prim^{3}} \} .
\end{equation}
The curve belongs to a bundle of nonintersecting curves
defined, for instance, by $\beta = \phi_{0}\alpha + \prim^{3}
\alpha^{2} + \prim^{5} \alpha^{4}$.

\subsubsection{$\alpha$-curve}

To the parametric curve
\begin{equation}
  \alpha = \prim^{2} \kappa^{4},
  \qquad \qquad
  \beta = \prim^{2} \kappa + \kappa^{2} + \prim \kappa^{4},
\end{equation}
it corresponds the matrices
\begin{equation}
\mathsf{W}_{\vec{\alpha}} = \left (
  \begin{array}{ccc}
    0 & 0 & \prim^{2} \\
    \prim^{4} & 0 & 0 \\
    0 & \prim & 0
  \end{array}
\right ) ,
\qquad \qquad
\mathsf{W}_{\vec{\beta}} = \left  (
  \begin{array}{ccc}
    \prim^{2} & 1 & \prim \\
    \prim^{2} & \prim^{4} & 1 \\
    1 & \prim^{4} & \prim
  \end{array}
\right ) .
\end{equation}
Now we have $\det \mathsf{W}_{\vec{\alpha}} = 1$ and $\det
\mathsf{W}_{\vec{\beta}}=0$, which leads to the following explicit
form of the $\alpha$-curve
\begin{equation}
  \beta =\prim^{6} \alpha + \prim^{5} \alpha^{2} + \prim^{6} \alpha^{4},
\end{equation}
whose coefficients satisfy again the condition (\ref{cc}).  The
corresponding $\mathsf{W}_{\vec{\phi}}$ matrix is degenerate in this
case. The set of commuting operators is
\begin{equation}
  \{ Z_{\prim^{6}}, \, Z_{\prim^{3}} X_{\prim^{2}}, \,
  Z_{\prim^{7}} X_{\prim^{5}}, \, Z_{\prim^{4}} X_{\prim^{2}}, \,
  Z_{\prim} X_{\prim^{3}}, \, Z_{\prim^{5}} X_{\prim^{3}}, \,
  Z_{\prim^{2}}X_{\prim^{5}} \},
\end{equation}
and the curve belongs to the bundle $\beta = \psi_{0}\alpha +
\prim^{5}\alpha^{2}+\prim^{6}\alpha^{4}$. 

\section{Exceptional curves}
\label{Sec: Exceptional}

The analysis of exceptional curves is considerably more involved.  
As we have stressed above, the points on the curve do not take all 
the values in the field and their admissible values are fixed by the
structural equations
\begin{equation}
  \sum_{m=0}^{r_{\alpha}} \upsilon_{m} \, \alpha^{2^{m}} = 0,
  \qquad \qquad
  \sum_{m=0}^{r_{\beta}} \tau_{m} \, \beta^{2^{m}} = 0,
  \label{sse}
\end{equation}
where $r_{\alpha}= \rank \mathsf{W}_{\vec{\alpha}} \leq n-1$ and
$r_{\beta} = \rank \mathsf{W}_{\vec{\beta}} \leq n-1$, which are a
consequence of the linear dependence of $\alpha^{2^{m}}$ and
$\beta^{2^{m}}$. The coordinates $\alpha $ and $\beta $ of an exceptional
curve are $\deg \alpha = 2^{n-r_{\alpha}}$ and $\deg \beta =
2^{n-r_{\beta}}$ times degenerate, respectively.  In other words, if
$(\alpha_{j},\beta_{j})$ is a point of an exceptional curve, for each
$\alpha_{j}$ there are $2^{n-r_{\beta}}$ values of $\beta$, such that
the points $(\alpha_{j},\beta_{k})$ ($k=1, \ldots , 2^{n-r_{\beta}}$)
belong to the same curve and, conversely, for each $\beta_{j}$ there
are $2^{n-r_{\alpha}}$ values of $\alpha $, such that the points
$(\alpha_{k},\beta_{j})$ also belong to the same curve.

Due to the nonsingularity condition, there are $2^{n}$ different 
pairs of points $(\alpha , \beta)$ belonging to the curve, so 
the condition $r_{\alpha} + r_{\beta} \geq n$ is satisfied. For instance,
for $\Gal{2^2}$ the only possibility is $r_{\alpha} = r_{\beta} = 1$,
and the only type of degeneration is $\deg \alpha = \deg \beta =2$.
For $\Gal{2^3}$ there are three possibilities: $r_{\alpha} = r_{\beta}
= 2$ ($\deg \alpha = \deg \beta =2$), $r_{\alpha} = 1, r_{\beta} = 2$
($\deg \alpha = 4$, $\deg \beta =2$), and $r_{\alpha} = 2, r_{\beta} =
1$ ($\deg \alpha = 2$, $\deg \beta = 4$).

When a curve equation can be found  [i.e., a relation of  the type
$F( \alpha ) = G (\beta )$, where $F ( \alpha )$ and $G ( \beta )$
are polynomials of degrees $2^{(r_{\alpha}-1)}$ and $2^{(r_{\beta}-1)}$], 
it  establishes  a direct correspondence between the roots of (\ref{sse}). 
To define uniquely the curve, the equation $F(\alpha) = G(\beta )$ 
should be supplemented with the structural equations. Nevertheless, 
such a relation exists only when the conditions $r_{\alpha, \beta} \geq 
(n+1)/2$ hold.

There are two ways of approaching the classification of exceptional
curves.  The first is a direct analysis of the parametric form
(\ref{curve1}) [whose coefficients satisfy the commutativity relation
(\ref{cc_g2}) and the corresponding determinants vanish]. We have to
determine the rank of the matrices $\mathsf{W}_{\vec{\alpha}}$ and
$\mathsf{W}_{\vec{\beta}}$, find the structural relations (\ref{sse}),
and check the nonsingularity condition. The main difficulty with this
approach is the complicated form of (\ref{cc_g2}), which is related to
that fact that there is no one-to-one correspondence between the
parametric form of a curve and points in the discrete phase space, in
the sense that the same curve can be defined by several different
parametric equations.

We shall take an alternative route and construct all the possible
exceptional curves by imposing \textit{ab initio} the nonsingularity
and commutativity conditions. An important ingredient in this
construction is the existence of \textit{a priori} information about
the degree of degeneration in the $\alpha $ and $\beta $ directions.
We shall outline the main idea and study in detail only the
eight-dimensional case.

\subsection{Constructing exceptional curves}

Let us consider a nonsingular curve with degenerations
$2^{n-r_{\alpha}}$ and $2^{n-r_{\beta}}$ along the $\alpha$ and
$\beta$ axes, respectively. The structural equations (\ref{sse}) 
can be represented as
\begin{equation}
  \alpha \, \prod_{j=1}^{2^{r_{\alpha}}-1} ( \alpha + \alpha_{j} ) =0,
  \qquad \qquad
  \beta \, \prod_{j=1}^{2^{r_{\beta}}-1} ( \beta + \beta_{j} ) =0,
  \label{se3}
\end{equation}
with all the roots $\alpha_j$ and $\beta_j$ different.  Since
only the powers $\alpha^{2^{m}}$ and $\beta^{2^{m}}$ ($m=0, \ldots,
r_{\alpha, \beta}$) can appear in (\ref{se3}), we obtain the
following restriction on the roots $\alpha_{j}$ and $\beta_{j}$
\begin{equation}
  \label{symfun}
  S_{k} (\alpha_{j} ) = 0,
  \qquad \qquad
  S_{k} ( \beta_{j} ) = 0,
  \qquad k \neq 2^{r_{\alpha, \beta}} - 2^m
  \label{sf1}
\end{equation}
where
\begin{equation}
  S_{k}( \xi ) = \sum_{i_{1} < i_{2} < \ldots < i_{k}}
  \xi_{i_{1}} \xi_{i_{2}} \ldots \xi_{i_{k}},
\label{S1}
\end{equation}
are symmetrical functions of the roots. This restriction implies that
only $r_{\alpha}$ ($r_{\beta}$) roots $\alpha_{j}$ ($\beta_{j}$) are linearly
independent.

Condition (\ref{Tr_cond}) implies that, given a degree of degeneration
and once one of the structural equations is fixed, the other
structural equation is not arbitrary. In other words, having
determined the admissible points along one of the axis, all the
admissible points of an additive commutative curve along the other 
axis are uniquely defined.

The simplest situation corresponds to the case when degeneration
along both axes is the same: $\deg \alpha = \deg \beta = g
=2^{n-r}$ ($r \geq n/2$), i.e., the ranks of the corresponding
matrices (\ref{MN}) are $r = r_{\alpha} = r_{\beta}$. Suppose that the
structural equation for $\alpha $ is fixed. Then, the commutativity
condition is equivalent to the following set of equations
\begin{equation}
  \tr (\alpha_{k} \beta_{j}) = 0,
  \qquad \qquad
  \left \{
  \begin{array}{ll}
    j = 1, \ldots , n-r, & \quad k = 1, \ldots, r,   \\
    j = 1, \ldots, r, &  \quad k=1, \ldots , n-r,
   \end{array}
   \right .
\label{com_exp_2_2}
\end{equation}
where $(\alpha_{k}, \beta_{k})$ are linearly independent roots and the points
$(\alpha_{j},0)$ and $(0,\beta_{j})$, with $j=1, \ldots ,2^{n-r}$, belong to
the curve.

Equations~(\ref{com_exp_2_2}) mean that (\ref{se3}) can be always
written as $\tr (\alpha \, \xi )=0, \tr (\beta \ \zeta )=0,$ with $\xi$ and 
$\zeta$ being fixed elements of $\Gal{2^n}$.

In the doubly degenerate case, $g=2$, the value of $\beta_{1}$ is uniquely
determined from the first condition in (\ref{com_exp_2_2}) and the
curve can be represented as a disjoint union of two straight lines
\begin{equation}
  \beta^{(1)} = \frac{\beta_{1}}{\alpha_{1}} \, \alpha ,
  \quad 
  \beta^{(2)} = \frac{\beta_{1}}{\alpha_{1}} \, \alpha + \beta_{1} ,
   \label{ec_g_d2}
\end{equation}
with $\alpha = \alpha_{1}, \ldots , \alpha_{2^{r}-1}$.  It is worth noting
that (\ref{com_exp_2_2}) in this case is just a structural equation, so
that
\begin{equation}
  \label{Gun}
  \beta_{1} = \frac{\upsilon_{1}}{\upsilon_{0}} =
  \frac{S_{2^{r}-2}}{S_{2^{r}-1}},
\end{equation}
where $\upsilon_{k}$ are the coefficients in (\ref{sse}) and $S_{r}$
are the symmetrical functions (\ref{symfun}) of arguments $\alpha_{k}$
($k = 1, \ldots, 2^{r}-1$).  Each different ordered set $\{\alpha_{1},
\ldots,\alpha_{r}\}$, with $\alpha _{j} \neq \alpha_{k} \neq 0$ determines 
thus an exceptional doubly degenerate curve.

For higher degenerations, the curve is a disjoint union of $g$
straight lines
\begin{equation}
  \beta^{(1)} = \lambda \alpha , \;
  \beta^{(2)} = \lambda \alpha + \beta_{1} , \ldots, \;
  \beta^{(g)} = \lambda \alpha + \beta_{g-1},
  \label{ec_g_d}
\end{equation}
where $\beta_{m}$ ($m = r + 1, \ldots , g-1$) are obtained as
the possible different linear combinations of $\beta_{j}$ ($j=1,
\ldots, r$) and
\begin{equation}
  \lambda =\frac{\beta_{1}}{\alpha_{1}}= \ldots =
  \frac{\beta_{r}}{\alpha_{r}} .
\end{equation}
The above ordering indicates that the points
$(\alpha_{j},\beta_{j})$ belong to the same straight line.  Then,
$\beta_{k} = \beta_{1}\alpha_{k}/\alpha_{1}$ ($k = 2, \ldots,r$) and
$\beta_{1}$ are uniquely expressed in terms of admissible values of
$\alpha_{j}$ from (\ref{com_exp_2_2}), which is convenient to
rewrite in terms of the parameter $\lambda$: $ \tr( \lambda
\alpha_{j} \, \alpha_{k}) = 0$, for $j, k= 1, \ldots ,r$. It is
clear that the exceptional curves constructed using the
equations~(\ref{ec_g_d2}) to (\ref{ec_g_d}) are nonsingular.

If the degeneration along $\alpha $ and $\beta$ axes are different
(say, $r_{\alpha}>r_{\beta}$), then the curves can be represented
as a collection of nonintersecting  ``parallel'' curves 
\begin{equation}
  \beta^{(1)} = f(\alpha ), \;
  \beta^{(2)} = f(\alpha ) + \delta_{1}, \ldots , \;
  \beta^{(2^{n-r_{\alpha}})} = f(\alpha )  +\delta_{2^{n-r_{\alpha}}-1},
   \label{b_exp}
\end{equation}
where $f(\alpha )$ is the function
\begin{equation}
f(\alpha )=\sum_{k=0}^{n-r_{\alpha}-1} f_{k} \, \alpha^{2^{k}},
\end{equation}
and  the commutativity condition leads to the restrictions
$\tr (\delta_{i}\alpha_{j})=0$, which fix the values of
$\delta_{i}$.

The intersection problem can be studied using the same criterion as 
for the regular curves, taking into account that those conditions 
should be satisfied only at the admissible points of the curve.

\subsection{Examples}

\subsubsection{Four-dimensional case}

In the case of $\Gal{2^2}$ the only exceptional curves are doubly
degenerate, $r_{\alpha} = r_{\beta} = 1$. Besides, the structural 
equation is of second order: $\alpha ( \alpha +\alpha_{1} ) = 0$, 
so that any one of the three possible exceptional curves can be 
represented as a union of straight lines:
\begin{equation}
  \beta^{(1)} = \alpha_{1}^{-2} \alpha ,
  \qquad \qquad
  \beta^{(2)}=\alpha_{1}^{-2} \alpha + \alpha_{1}^{-1},
  \label{ec_d_4_1}
\end{equation}
with $\alpha = 0, \alpha_{1}$. More explicitly, for a given value 
of $\alpha_{1}$  we have the following curve:
\begin{equation}
(0,0), (\alpha_{1}, \alpha_{1}^{-1}), (0, \alpha_{1}^{-1}),
(\alpha_{1}, 0).
\end{equation}
In this case it is impossible to write down an equation that
relates $\alpha $ and $\beta$.

As an example, consider the curve
\begin{equation}
(0,0), (\prim , \prim^{2}), (0, \prim^{2}), ( \prim ,0),
\end{equation}
where $\alpha_{1}= \prim$. The structural equations are then
\begin{equation}
\alpha^{2} = \prim \alpha ,
\qquad \qquad
\beta^{2} = \prim^{2}\beta .
\end{equation}
The implication of the curve type on the factorization of the
basis will be discussed in section~\ref{Sec: Factorization}.

\subsubsection{Eight-dimensional case}

Two types of exceptional curves exist in the case of $\Gal{2^3}$: (i)
doubly degenerate in both directions, which corresponds to
$r_{\alpha}=r_{\beta}=2$; (ii) doubly degenerate in one direction and
quadruply degenerate in the other, which corresponds to $r_{\alpha}=2,
r_{\beta}=1$ or $r_{\alpha}=1, r_{\beta}=2$.

In the case (i), any exceptional curve can be represented as a union of
two lines
\begin{equation}
  \beta^{(1)} = \beta_{1}\alpha_{1}^{-1} \alpha ,
  \quad \qquad
  \beta^{(2)} = \beta_{1} ( \alpha_{1}^{-1} \alpha +1 ) ,
\end{equation}
where, according to equation~(\ref{Gun}),
\begin{equation}
  \beta_{1} = \frac{1}{\alpha_{1} + \alpha_{2}} +
  \frac{1}{\alpha_{1}}+\frac{1}{\alpha_{2}},  
  \label{beta_1}
\end{equation}
and the admissible values of $\alpha $ are $0, \alpha_{1},
\alpha_{2}$, and $\alpha_{1} + \alpha_{2}$. For a fixed set
$\{\alpha_{1},\alpha_{2}\}$ the following exceptional curve 
is defined
\begin{eqnarray}
  \label{e_c__8_dd}
  &  (0,0), (\alpha_{1},0), (\alpha_{1},\beta_{1}),
  (\alpha_{2},\beta_{1} ( \alpha_{1}^{-1} \alpha_{2} + 1)),
  (\alpha_{2}, \beta_{1}\alpha_{1}^{-1}\alpha_{2}), & \nonumber  \\
  & (\alpha_{1} + \alpha_{2},\beta_{1}\alpha_{1}^{-1} \alpha_{2}),
  (\alpha_{1} + \alpha_{2}, \beta_{1}( \alpha_{1}^{-1}\alpha_{2}+1 )),
  ( 0,\beta_{1})  . &
\end{eqnarray}
From the above equation we find  that there are 21
exceptional curves due to the permutational symmetry between
$\alpha_{2}$ and $\alpha_{1} + \alpha_{2}$.

As an example, consider
\begin{equation}
(0,0), (\prim^{4}, 0), (\prim^{4}, \prim^{5}),
(\prim^{3},\prim^{7}), (\prim^{3}, \prim^{4}),
(\prim^{6},\prim^{4}),(\prim^{6}, \prim^{7}),
(0,\prim^{5}),
\end{equation}
where $\alpha_{1} = \prim^{4}, \alpha_{2} = \prim^{3}$. The
structural equations are
\begin{equation}
  \prim^{6}\alpha + \prim^{4}\alpha^{2} + \alpha^{4} = 0,
  \qquad \qquad
  \prim^{2}\beta +\prim^{6} \beta^{2}+\beta^{4}=0,
\end{equation}
and the curve has the form
\begin{equation}
  \beta^{2} + \prim^{5} \beta = \prim^{6} \alpha +
  \prim^{2}\alpha^{2} .
\end{equation}

In the case (ii), one of coordinates (say, for instance, $\alpha $)
is still doubly degenerate, while the other one is quadruply degenerate.
Then, the coordinate $\beta$  takes only two values: $0$ and $\delta$,
while the allowed values of $\alpha$ are $0, \alpha_{1}, \alpha_{2}$, 
and $\alpha_{1} + \alpha_{2}$, so such a curve has the form
\begin{equation}
\beta^{(1)} = f (\alpha ),
\qquad \qquad
\beta^{(2)} = f(\alpha ) + \delta ,
\end{equation}
where
\begin{equation}
f (\alpha )= \frac{\delta}{\alpha_{2}(\alpha_{1}+\alpha_{2})}
( \alpha_{1}\alpha + \alpha^{2} ) ,
\end{equation}
and $\delta $ satisfies $\tr (\delta \, \alpha_{1,2})=0$, which leads 
to $\delta = \beta_{1}$.

Explicitly, the points of such a curve are
\begin{equation}
  ( 0,0), (\alpha_{1}, 0), (\alpha_{2}, 0),
  (\alpha_{1} + \alpha_{2}, 0),(\alpha_{1}, \delta ),
  (\alpha_{2}, \delta ),( \alpha_{1}+\alpha_{2},\delta ),
  ( 0,\delta ) ,  
  \label{sec_8}
\end{equation}
so there are seven different curves of this type due to permutational
symmetry between $\alpha_{1},\alpha_{2}$, and $\alpha_{1}+\alpha_{2}$.

As an example of such a degenerate curve, consider the points
\begin{equation}
  (0,0), (\prim^{3}, 0), ( \prim^{5}, 0),
  ( \prim^{2}, 0), (\prim^{3}, \prim^{6}),
  (\prim^{5},\prim^{6}),(\prim^{2},\prim^{6}),
  (0,\prim^{6}),
\end{equation}
where $\alpha_{1}=\prim^{3},\alpha_{2}=\prim^{5}$. The
corresponding structural equations are
\begin{equation}
  \prim^{3} \alpha + \prim^{2} \alpha^{2}+ \alpha^{4} = 0,
  \qquad \qquad
  \prim^{6} \beta +\beta^{2} = 0.
\end{equation}

\section{Local transformations}

Local transformations induce nontrivial transformations in the curve,
although they preserve the factorization properties (in a given
basis). We recall that in any selfdual basis we can represent the
monomial $Z_{\alpha} X_{\beta}$ in the following way
\begin{eqnarray}
  Z_{\alpha} X_{\beta} = \otimes \prod_{j=1}^{n}
  \sigma_{z}^{a_{j}} \sigma_{x}^{b_{j}} \equiv
  \otimes \prod_{j=1}^{n} ( a_{j},b_{j}) ,
\end{eqnarray}
where $\otimes \prod_j$ denotes the tensor product over the index $j$.
Under local transformations (rotations of angle $\pi/2$ around $z, x$,
and $y$ axes, which we call $z$-, $x$- and $y$-rotations) applied to
the $j$th particle, it transforms as
\begin{eqnarray}
  z &:& ( a_{j}, b_{j} ) \mapsto ( a_{j} + b_{j}, b_{j}) ,
  \label{LT_z} \nonumber \\
  x &:& ( a_{j}, b_{j} ) \mapsto ( a_{j}, b_{j} + a_{j}) ,
  \label{LT_x} \\
  y &:& ( a_{j}, b_{j} ) \mapsto (a_{j}+ a_{j}+ b_{j},
  b_{j} + a_{j} + b_{j}) = ( b_{j},a_{j} ) .
  \label{LT_y} \nonumber
\end{eqnarray}
To give a concrete example, suppose we consider a $z$-rotation. The
operator $\sigma_{z}$, corresponding to $(a_{j}=1,b_{j}=0)$, is
transformed into $(a_{j} = 1 + 0 = 1, b_{j}=0)$; i.e., into itself,
while the operator $\sigma_{x}$, corresponding to $(a_{j}=0,b_{j}=1)$,
is mapped onto $(a_{j}=0+1=1,b_{j}=1)$, which coincides with
$\sigma_{y}$. In the same way $\sigma_{y}$ is mapped onto
$\sigma_{x}$, while the identity ($a_{j}=0, b_{j}=0$) is mapped onto
itself.

In terms of the field elements it is equivalent to
\begin{eqnarray}
  z &:& \left \{
    \begin{array}{l}
      \displaystyle
      \alpha \mapsto \alpha^{\prime} = \alpha +
      \sum_{k} \basis_{k} \tr ( \beta \basis_{k}) , \\
      \beta \mapsto \beta^{\prime} = \beta ,
    \end{array}
  \right . \nonumber \\
  x &:& \left \{
    \begin{array}{l}
      \alpha \mapsto \alpha^{\prime} = \alpha ,  \\
      \displaystyle
      \beta \mapsto \beta^{\prime} = \beta +
      \sum_{k}\basis_{k}\tr ( \alpha \basis_{k})  ,
    \end{array}
  \right . \\
  y &:& \left \{
    \begin{array}{l}
      \displaystyle
      \alpha \mapsto \alpha^{\prime} = \alpha  +
      \sum_{k}\basis_{k}\tr [ (\alpha + \beta ) \basis_{k} ]  ,  \\
      \displaystyle
      \beta \mapsto \beta^{\prime} = \beta +
      \sum_{k}\basis_{k}\tr [ ( \alpha +\beta ) \basis_{k} ]  ,
    \end{array}
  \right . \nonumber
\end{eqnarray}
where ${\basis}$ is the selfdual basis.

These transformations are nonlinear in the field elements: starting
with a standard set of MUB operators related to rays, we obtain
another set of MUB operators parametrized with points of curves, but
leading to the same factorization structure. Indeed, consider a ray as
in equation~(\ref{ray_pf}). Then, under $z$-, $x$-, and $y$-rotations
we have
\begin{eqnarray}
  z &:&
  \left \{
    \begin{array}{l}
      \displaystyle
      \alpha \mapsto \alpha^{\prime} =  \mu \kappa +
      \sum_{m=0}^{n-1} \kappa^{2^{m}} \nu^{2^{m}}
      \sum_{k} \basis_{k}^{2^{m}+1},  \\
      \displaystyle
      \beta \mapsto \beta^{\prime} = \nu \kappa ,
    \end{array}
  \right . \nonumber \\
  x &:&  \left \{
    \begin{array}{l}
      \alpha \mapsto \alpha^{\prime} =  \mu \kappa ,  \\
      \displaystyle
      \beta \mapsto \beta^{\prime} = \nu \kappa  +
      \sum_{m=0}^{n-1}\kappa^{2^{m}} \mu^{2^{m}}
      \sum_{k} \basis_{k}^{2^{m}+1},
    \end{array}
  \right . \\
  y &:& \left \{
    \begin{array}{c}
      \displaystyle
      \alpha \mapsto \alpha^{\prime} = \mu \kappa +
      \sum_{m=0}^{n-1} \kappa^{2^{m}} (\nu + \mu )^{2^{m}}
      \sum_{k} \basis_{k}^{2^{m}+1} \\
      \displaystyle
      \beta \mapsto \beta^{\prime} = \nu \kappa +
      \sum_{m=0}^{n-1}\kappa^{2^{m}} (\nu + \mu )^{2^{m}}
      \sum_{k}\basis_{k}^{2^{m}+1}
    \end{array}
  \right . \nonumber
\end{eqnarray}

Note that the $z$- and $x$-rotations produce regular curves
\begin{eqnarray}
  z &:&\alpha = \mu \nu^{-1} \beta \mapsto \alpha^{\prime}=
  \mu \nu^{-1} \beta + \sum_{m=0}^{n-1}\beta^{2^{m}}
  \sum_{k}\basis_{k}^{2^{m}+1},  \nonumber \\
  & & \\
  x &:&\beta = \mu^{-1}\nu \alpha  \mapsto \beta^{\prime}=
  \mu^{-1}\nu \alpha +\sum_{m=0}^{n-1}\alpha^{2^{m}}
  \sum_{k}\basis_{k}^{2^{m}+1} . \nonumber
\end{eqnarray}
Meanwhile, the $y$-rotation may lead to exceptional curves. In this
case we always have $\kappa = ( \mu + \nu )^{-1} (\alpha +\beta ) $,
and thus the explicit equation of that curve is either
\begin{equation}
  \alpha =\mu ( \nu + \mu )^{-1}  ( \alpha +\beta ) +
  \sum_{m=0}^{n-1} ( \alpha +\beta )^{2^{m}}
  \sum_{k}\basis_{k}^{2^{m}+1},
\end{equation}
or
\begin{equation}
  \beta = \nu ( \nu + \mu )^{-1} ( \alpha + \beta)
  + \sum_{m=0}^{n-1} ( \alpha +\beta )^{2^{m}}
  \sum_{k} \basis_{k}^{2^{m}+1}.
\end{equation}

For instance, in the two-qubit case, starting from the ray $\beta =0$,
we can generate all the curves shown in table~\ref{table1}.  In
particular, it can be proven that in there are only two equivalence
classes of curves\cite{Klimov2007}.

\begin{table}
  \caption{\label{table1}
    Curves generated by applying the rotations indicated in the left column
    to the ray $\beta = 0$ in the case of two qubits.}
  \begin{center}
    \begin{tabular}{ll}
      \hline 
      Rotation \qquad \qquad \qquad & Curves  \\
      \hline
      $ x \otimes \openone$ & 
      $\beta =\prim \alpha + \alpha^{2}, \, 
      \beta^{2} = \prim^{2}\beta$  \\ 
      $ y \otimes \openone$ & 
      $\prim \beta + \beta^{2} = \prim^{2} \alpha +\alpha ^{2}, \,
      \alpha^{2} = \prim^{2} \alpha , \,
      \beta^{2} = \prim \beta $ \\ 
      $ \openone \otimes x$ & 
      $\beta = \prim^{2} \alpha + \alpha^{2}, \,
      \beta^{2} = \prim \beta$ \\ 
      $\openone \otimes y $ & 
      $\prim^{2} \beta + \beta^{2} = \prim \alpha + \alpha^{2}, \,
      \alpha^{2} = \prim \alpha , \,
      \beta^{2} = \prim^{2}\beta $ \\ 
      $x \otimes x$ & 
      $\beta = \alpha$  \\ 
      $y \otimes y$ & 
      $\beta = \prim^{2} \alpha + \alpha^{2}, \,
      \beta ^{2}=\prim \beta$ \\ 
      $x\otimes y $ & 
      $\prim \beta + \prim^{2} \beta^{2} = \alpha + 
      \prim ^{2}\alpha^{2}, \,
      \alpha^{2} = \prim \alpha , \,
      \beta^{2}=\prim \beta $  \\ 
      $y \otimes x$ & 
      $\alpha = \prim \beta + \beta^{2},\,
      \alpha^{2} = \prim^{2} \alpha$ \\
      \hline
    \end{tabular}
  \end{center}
\end{table}

\section{Factorization structure and curve bundles}
\label{Sec: Factorization}

In this section we discuss bundles leading to MUBs with different
factorization structures. As we have stated before, given a basis in
the field, any operator, labeled by a point of a curve, is factorized
into a product of one-particle Pauli operators. For qubit systems the
selfdual is the most appropriate, for the Fourier operator is
factorized, and thus, the factorization of $Z_{\alpha}$ and
$X_{\alpha}$ is straightforward. Now, let us divide each monomial
$Z_{\alpha}X_{\beta}$, into two parts, so that the first part contains
$k$ Pauli operators and the second part $n-k$ operators. If any first
``block'' of the set of $d-1$ commuting generalized Pauli operators
commutes with all the other ``blocks'', we will say that the
corresponding curve is factorized into two sets. Obviously, the second
blocks would then also commute between them. Moreover, inside the
first or second blocks may exist some ``sub-blocks'' that commute with
corresponding sub-blocks, etc. In the end, we can represent any curve
$\Gamma \in \Gal{2^n}$ in the following factorized form:
\begin{equation}
  \Gamma =\{ m_{1},m_{2}, \ldots ,m_{N}\},
  \qquad \qquad
 0 < m_{1}\leq m_{2}\leq \ldots
\leq  m_{N},
\;
\sum_{i}m_{i}=n,  \label{1_curve_part}
\end{equation}
where $m_{i}\in \mathbb{N}$ is the number of particles in the 
$i$-th block that cannot be factorized anymore. It is clear that
$\{m_{1},m_{2},\ldots, m_{N}\}$ is just a partition of the integer
$n$, so the maximum number of terms is $n$, which corresponds to a
completely factorized curve, $\Gamma =\underbrace{\{1,1, \ldots,1\}}_{n}$, 
and the minimum number of terms is one, corresponding to a completely 
nonfactorized curve, $\Gamma = \{ n \}$.

One can construct bundles that contain only regular curves, as it was
shown in section~\ref{Sec: Regular}. A systematic construction of
bundles containing both regular and exceptional curves is a more
involved task, which can be carried out numerically for low
dimensions.

As an example, consider the ray $\beta = 0$ over $\Gal{2^2}$, so 
the corresponding set of operators is $\{Z_{\prim},Z_{\prim^{2}},
Z_{\prim^{3}} \}$. In the selfdual basis $(\prim ,\prim^{2})$ this
set is factorized into $( \sigma_{z} \openone, \openone \sigma_{z},
\sigma_{z}\sigma_{z})$. Then, the curve $\beta =0$ is represented as
$\Gamma = \{1, 1\}$, i.e., both particles are factorized. The ray
$\beta = \prim \alpha $, whose points label the set $\{ Z_{\prim}
X_{\prim^{2}}, Z_{\prim^{2}} X_{\prim^{3}}, Z_{\prim^{3}} X_{\prim}
\}$, has the following factorization $( \sigma_{z}\sigma_{x},
\sigma_{x}\sigma_{y}, \sigma_{y} \sigma_{z}) $, so that it can be
represented as $\Gamma =\{ 2 \}$, which means that there are no
factorized blocks. In the case of three qubits the possible partitions
are $\{1, 1, 1\}$ (e.g., the ray $\beta =0$), $\{1, 2 \}$ (e.g., the
regular curve $\beta =\prim^{6} \alpha^{2} + \prim^{3} \alpha^{4}$),
and $\{ 3 \}$ (e.g., the ray $\beta =\prim^{3} \alpha$).

The representation (\ref{1_curve_part}) is invariant under local
transformations.  The corresponding basis preserves the factorization
of the operator set, and local transformations preserve the
factorization structure of the curve.  This means that all the
completely factorized curves can be obtained from a single factorized
ray, say $\beta =0$.  Nevertheless, the curves with the same
factorization structure are not necessarily equivalent under local
transformations (except in the trivial two-qubit
case~\cite{Klimov2007}).

A bundle may contain curves with different factorizations. We can
characterize different bundles with a set of numbers that indicate 
the number of completely factorized curves ($\underbrace{\{1,1, 
\ldots,1\}}_{n}$ structure), completely factorized except a single
two-particle block (curve of the type $\{\underbrace{1,1,\ldots,
1}_{n-2},2\}$), etc., until completely nonfactorized curves $\{n\}$. 
In other words, we assign to the bundle the set of numbers
\begin{equation}
  (k_{1},k_{2}, \ldots ,k_{p(n)}),
  \qquad  \qquad
  \sum_{j}k_{j}=2^{n}+1,
\end{equation}
which indicate the number of curves factorized in $n$ one-dimensional
blocks, $k_{1}$; the number of curves factorized in $n-2$
one-dimensional blocks and one two-dimensional block, $k_{2}$, etc,
and $p(n)$ is the number of partitions of an integer $n$.

\subsection{Curves over $\Gal{2^2}$}

As we have discussed, an additive commutative curve over $\Gal{2^2}$
can be expressed as
\begin{equation}
  \alpha (\kappa) = \alpha_{0} \kappa + \alpha_{1} \kappa^{2},
  \qquad \qquad
  \beta (\kappa) = \beta_{0} \kappa + \beta_{1}\kappa^{2},
\end{equation}
where the commutativity condition (\ref{cc1}) impose the following
restrictions on the coefficients $\alpha_{j}$ and $\beta_{j}$
\begin{equation}
  \alpha_{1} \beta _{0}+ ( \alpha_{1} \beta_{0})^{2} = 
  \alpha_{0} \beta_{1} + (\alpha_{0} \beta_{1})^{2}.  
  \label{Eq:condition}
\end{equation}
In this simple case, the whole analysis can be carried out from the
parametric form. Nevertheless, it is more convenient to separate types
of curves on regular and exceptional, according to our discussion in
sections~\ref{Sec: Regular} and \ref{Sec: Exceptional}.

All the possible additive commutative structures can be divided into
two types:

a) 12 regular curves, which can be constructed according to the
general rule (\ref{cc}), among which there are four rays
\begin{equation}
  \beta = \lambda \alpha ,
  \qquad  \qquad
  \alpha =0,  \label{rays_4}
\end{equation}
and 8 curves
\begin{equation}
  \label{rc_4}
  \alpha\mathrm{-curves}: \; \beta = \eta \alpha + \alpha^{2},
  \qquad  \qquad
  \beta\mathrm{-curves}: \; \alpha =\eta \beta + \beta^{2}.
\end{equation}

b) 3 exceptional curves that can be represented as a union of two
parallel lines (\ref{ec_d_4_1}) or in the parametric form
\begin{equation}
  \alpha = \mu (\kappa +\kappa^{2}) ,
  \qquad
  \beta =\mu^{2} (\prim \kappa + \prim^{2} \kappa^{2}) .
  \label{exep_4} 
\end{equation}
Every point of these exceptional curves is doubly degenerate and the
admissible values of $\alpha $ and $\beta $ are $\{0,\mu \}$ and
$\{0,\mu^{2}\}$, respectively.

It is important to stress that it is possible to obtain all the curves
of form (\ref{rays_4}), (\ref{rc_4}), and (\ref{exep_4}) from the rays
after some (nonlinear) operations, corresponding to local
transformations of operators~\cite{Klimov2007}. The families of such
transformations are the following: 8 curves (the rays $\beta
=\alpha $ and $\beta =0$ among them) can be obtained from the single
ray $\alpha =0$ (corresponding to the vertical axis) and the other
5 curves (the ray $\beta =\prim^{2}\alpha $ among them) from the
ray $\beta = \prim \alpha $.

The simplest curve bundle contains just rays. There are three rays
($\beta = \alpha $, $\beta = 0$, and $\alpha =0$) with completely
factorizable structure $\{1, 1\}$ and two rays ($\beta = \prim \alpha$
and $\beta = \prim^{2}\alpha$ with nonfactorizable structure $\{ 2
\}$.  Since any other bundle can be obtained from the ray bundle by
applying some local transformations, the only bundle structure is
$(3,2)$, i.e., in any bundle there are three factorizable curves and
two nonfactorizable (having EPR-states as basis states).

\subsection{Curves over $\Gal{2^3}$}

A generic additive commutative curve over $\Gal{2^3}$ is given by
\begin{equation}
  \alpha = \alpha_{0}\kappa + \alpha_{1}\kappa^{2} + \alpha_{2}\kappa^{4},
  \qquad \qquad
  \beta = \beta_{0}\kappa + \beta_{1} \kappa^{2} + \beta_{2} \kappa^{4}.
  \label{8_curve_1}
\end{equation}
The commutative condition in this case is much more complicated 
than for $\Gal{2^2}$, so a full analysis of all the possible 
curves becomes cumbersome if we start with (\ref{8_curve_1}). Instead, 
we can follow the procedure of sections~\ref{Sec: Regular} and 
\ref{Sec: Exceptional}.

A generic regular curves has always one of the following forms
\begin{eqnarray}
  \beta = \phi_{0} \alpha + \phi^{2} \alpha^{2} + \phi \alpha^{4},  
  \qquad \qquad
  \alpha & = & \psi_{0} \beta + \psi^{2} \beta^{2}+ \psi \beta^{4} .
  \label{8d_c} 
\end{eqnarray}
for $\alpha$- and $\beta$-curves, respectively. All in all,
we get 100 different regular curves.  There are 21 
doubly degenerate exceptional curves of the form (\ref{e_c__8_dd}) 
and 14 exceptional curves (\ref{sec_8}), which are quadruply 
degenerate in one direction and doubly degenerate in the other.

The simplest way of forming bundles of commutative curves is given in
(\ref{b1}). Then, four bundles of nine curves each
\begin{equation}
  \beta = \phi_{0} \alpha + \phi^{2} \alpha^{2} + \phi \alpha^{4},
  \qquad \qquad
  \alpha  = 0,
\end{equation}
where $\phi_{0}\in \Gal{2^3}$ and $\tr (\phi) =0$, have the 
factorization structure $(3,0,6)$. The choice $\phi =0$ leads
to the ray structure (\ref{rays}). Among them, only three rays
($\beta =0, \alpha =0,\beta =\alpha$) have the structure $\{1,1,1\}$, 
while the other six rays have the structure $\{3\}$.

All the other bundles with $\tr (\phi) =0$ (i.e., $\phi$ takes the
values $\prim$, $\prim^{2}$, and $\prim^{4}$) can be generated from
the bundle with $\phi =0$ by applying local transformations.  In
particular, $\phi = \prim$ is generated by an $x$-rotation of the
first qubit, $\phi = \prim^{2}$ by an $x$-rotation of the second
qubit, and $\phi = \prim^{4}$ by an $x$-rotation of the first and
second qubits. 

Another four bundles of nine curves
\begin{equation}
  \beta =\phi_{0}\alpha +\phi^{2}\alpha^{2}+\phi \alpha^{4},
  \qquad \qquad
  \alpha = 0,
\end{equation}
where $\phi_{0}\in \Gal{2^3}$ and $\tr ( \phi ) =1$ generate all the
structures $(1,6,2)$. For instance, in the bundle with $\phi =1$ 
the curves with $\phi_{0}=0,\prim^{7}=1$ have the factorization
 $\{3\}$ and all the other values of $\phi_{0}\in
\Gal{2^3}$ generate the curves with the factorization $\{1,2\}$.  The
ray $\alpha =0$ has the factorization $\{1,1,1\}$.

As in the previous case, all the bundles with $\tr (\phi) =1$ ($\phi =
\prim^{3}, \prim^{5}, \prim^{6}$) can be obtained form the bundle with
$\phi =1$ by some local transformations: $\phi =\prim^{3}$ by an
$x$-rotation of the first qubit, $\phi = \prim^{6}$ by an $x$-rotation
of the second qubit and $ \phi =\prim^{5}$ by an $x$-rotation of the
third qubits.

The aforementioned local rotations can be also viewed as a
relabeling of the points of the curves $\beta =\phi_{0} \alpha
+\alpha^{2} + \alpha^{4}$ according to $\alpha \mapsto
\prim^{k} \alpha$ and $\beta \mapsto \prim^{-k}\beta $, with
$k=2,4,1$, respectively. On the other hand, $k=3,5,6$ correspond to
nonlocal transformations and lead to the bundles of curves
corresponding to the $(3,0,6)$ structure.

It is possible to obtain one more type of bundles with different
factorization structure and constituted only by regular curves. To
this end we recall that the nonintersection condition between two
regular curves has the form
\begin{equation}
  ( \phi_{0} + \phi_{0}^{\prime})^{7} +
  \tr [ ( \phi_{0}+\phi_{0}^{\prime} ) ( \phi + \phi^{\prime}) ] =1 ,
  \label{8d_ni}
\end{equation}
so $\phi_{0}$ and $\phi_{0}^{\prime}$ never coincide. Now, let
us take three nonintersecting regular curves satisfying
(\ref{8d_ni})
\begin{eqnarray}
  \tr [ ( \phi_{0} + \phi_{0}^{\prime})
  ( \phi + \phi^{\prime} ) ] & = & 0, \nonumber \\
  \tr [ ( \phi_{0}^{\prime} + \phi_{0}^{\prime \prime})
  ( \phi^{\prime}+\phi^{\prime \prime} ) ] & = & 0, \\
  \tr [ ( \phi_{0} + \phi_{0}^{\prime \prime})
  ( \phi + \phi^{\prime \prime} ) ] & = & 0, \nonumber
\end{eqnarray}
and construct a set of curves according to
\begin{equation}
  \beta = ( \phi_{0}^{a} + \phi_{0}^{b}) \alpha +
  ( \phi^{a} + \phi^{b} )^{2} \alpha^{2} +
  ( \phi^{a} + \phi^{b}) \alpha^{4},
\end{equation}
where $\phi_{0}^{a}$ , $\phi_{0}^{b}$ and $\phi^{a}$, $\phi^{b}$ are
coefficients of the previously defined curves. In this way we
generate five additional new curves:
\begin{eqnarray}
\label{8_3}
  \beta & = & ( \phi_{0}+\phi_{0}^{\prime} ) \alpha +
  ( \phi + \phi^{\prime} )^{2}\alpha^{2} +
  ( \phi +\phi^{\prime}) \alpha^{4},  \nonumber \\
  \beta & = & ( \phi_{0}+\phi_{0}^{\prime \prime} ) \alpha +
  ( \phi +\phi^{\prime \prime})^{2} \alpha^{2} +
  ( \phi +\phi^{\prime \prime}) \alpha^{4}, \nonumber  \\
  \beta &=& ( \phi_{0}^{\prime}+\phi_{0}^{\prime \prime}) \alpha +
  ( \phi^{\prime}+\phi^{\prime \prime} )^{2}\alpha^{2} +
  ( \phi^{\prime}+\phi^{\prime \prime}) \alpha^{4},  \\
  \beta &=& ( \phi_{0}+\phi_{0}^{\prime} + \phi_{0}^{\prime \prime} ) \alpha +
  ( \phi +\phi^{\prime}+\phi^{\prime \prime})^{2}\alpha^{2} +
  ( \phi +\phi^{\prime}+\phi^{\prime \prime}) \alpha^{4}, \nonumber  \\
  \beta &=&0,\nonumber
\end{eqnarray}
and one has to add the last curve $\alpha =0$ to complete the set of nine
curves.

We observe that the three ``initial'' curves can be obtained in the
same way using some of the curves in (\ref{8_3}). This implies that any curve
constructed in this way should satisfy the condition
\begin{equation}
  \tr (\phi_{0} \, \phi ) =0.
\end{equation}
All these sets of curves lead to the factorization structure
$(2,3,4)$. One example of such type of bundle is
\begin{equation}
\begin{array}{lcl}
  \{ 1,1,1 \} & \mapsto & \alpha =0,\beta =0 \\
  \{ 1,2 \} &  \mapsto & \left\{
    \begin{array}{l}
      \beta =\prim^{6}\alpha +\prim^{3}\alpha^{2}+\prim^{5}\alpha^{4}, \\
      \beta =\prim^{2}\alpha +\prim^{5}\alpha^{2}+\prim^{6}\alpha^{4}, \\
      \beta =\prim^{4}\alpha +\prim^{3}\alpha^{2}+\prim^{5}\alpha^{4}
     \end{array}
     \right . \\
  \{ 3 \} & \mapsto & \left\{
    \begin{array}{l}
      \beta  = \prim^{3}\alpha ,\\
      \beta = \prim^{5}\alpha +\prim^{5} \alpha^{2} + \prim^{6}\alpha^{4}, \\
      \beta = \prim \alpha +\prim^{2}\alpha^{2} + \prim \alpha^{4}, \\
      \beta = \alpha +\prim^{2} \alpha^{2} + \prim \alpha^{4}.
      \end{array}
      \right .
\end{array}
\end{equation}

There is one more type of bundles with the structure $(0,9,0)$, i.e.,
which contains only curves with the factorization $\{1,2\}$. Such
bundles always contain exceptional curves. One example of those
bundles is

a) Regular curves
\begin{equation}
\begin{array}{ll}
  \alpha = \prim^{2}\beta +\prim^{3}\beta^{2}+\prim^{5}\beta^{4},
  \qquad \qquad &
  \alpha = \prim^{6}\beta +\prim^{3}\beta^{2}+\prim^{5}\beta^{4}, \\
  \beta =\prim^{2}\alpha +\prim^{3}\alpha^{2}+\prim^{5}\alpha^{4},
   \qquad \qquad &
  \beta = \prim^{6}\alpha^{2}+\prim^{3}\alpha^{4}, \\
  \alpha =\beta +\prim^{6}\beta^{2}+\prim^{3}\beta^{4},
   \qquad \qquad &
  \beta =\alpha +\prim^{3}\alpha^{2}+\prim^{5}\alpha^{4}, \\
  \alpha =\prim^{3}\beta^{2}+\prim^{5}\beta^{4},
   \qquad \qquad &
\end{array}
\end{equation}
b) Exceptional curves
\begin{eqnarray}
  \beta^{2} + \prim^{5} \beta  &=& \prim^{2} \alpha^{2} + \prim^{6} \alpha ,
  \qquad \qquad 
  \tr ( \prim^{4}\beta ) =0, \quad \tr ( \prim^{5}\alpha ) =0; \nonumber \\
  & & \\
  \beta^{2}+\prim^{2}\beta  &=&\prim^{6}\alpha^{2}+\prim^{5}\alpha ,
  \qquad \qquad 
  \tr ( \prim^{6}\beta ) =0,\quad \tr (\prim^{2}\alpha ) =0, \nonumber
\end{eqnarray}
where the structural equations are written in the trace form.

The rays with $\{1,2\}$ factorization structure cannot be obtained
from rays by local transformations. Moreover, not all the curves with
factorization $\{ 3 \}$ can be obtained from the rays of the same
type.

All these structures can be obtained form each other by nonlocal
transformations, which can be always reduced to a combination of CNOT
gates and local transformations.  To each bundle with a given
factorization structure corresponds a set of nonlocal transformations
preserving such structure. Such a symmetry can be used to determine
the optimum tomographic procedure and will be discussed elsewhere.

The phase-space approach presented here also provides an alternative
to the graph-state classification of all the possible stabilizers
states for $n$-qubit systems. In fact, each additive curve
represents a basis in the $2^{n}$ dimensional Hilbert space, so that
the stabilizer state is one element of such basis. Not each curve can
be directly associated with a graph state~\cite{Raussendorf2003}, but
it can be reduced to an appropriate graph state through local Clifford
transformations~\cite{vandenNest2004}. While the classification of the
graph states represents a formidable task for large numbers of qubits,
the phase-space approach allows working with algebraic structures. 
Although the local equivalence is still an open problem, at least 
we can determine some elements of equivalence classes under local 
Clifford transformations in a relatively simple form. Besides,
several nonlocal qubit operations as SWAP or CNOT gates can be 
nicely represented in terms of curves transformations curves.  The 
above-mentioned problems will be analyzed in future work.

\section{Conclusions}

It has relatively recently been realized that several different types
of MUBs, with respect to their factorization properties, exist for a
system of $n$ qubits. Such bases are related to different arrangements
of generalized Pauli monomials into sets of commuting
operators. The construction of a whole set of MUBs is an involved
problem, especially for large number of qubits. The simplest MUB
construction was discovered by Wootters and it is related to straight
lines in the discrete phase space. We have shown that all the other
MUBs are connected with a special type of discrete curves.  Although,
in principle, we can classify all the possible additive commutative
curves and even determine which are related through local
transformations, arranging them in bundles of nonintersecting
curves is still an involved problem. Nevertheless, we can find some of
such bundles according to a ``recipe'' [specifically equations
(\ref{b1}) and (\ref{a1})], which represents an essential progress in
this field.

\ack 
The authors would like to acknowledge fruitful discussions with
Dr.  Iulia Ghiu, from University of Bucharest.  
This work was supported by the Grant No 45704 of Consejo Nacional de
Ciencia y Tecnolog\'{\i}a (CONACyT), the Swedish Foundation for
International Cooperation in Research and Higher Education
(STINT), the Swedish Research Council (VR), the Swedish Foundation
for Strategic Research (SSF), and the Spanish Research Directorate
(Grant FIS2005-06714).  A. B. K. was also supported by the Spanish 
Sabbatical Program (Grant SAB2006-0064).

\appendix

\section{Galois fields}
\label{Sec: Galois}

In this appendix we briefly recall the minimum background needed in
this paper.  The reader interested in more mathematical details is
referred, e.g., to the excellent monograph by Lidl and
Niederreiter~\cite{Lidl1986}.

A commutative ring is a set $R$ equipped with two binary operations,
called addition and multiplication, such that it is an Abelian group
with respect the addition, and the multiplication is associative. 
Perhaps, the motivating example is the ring of integers $\mathbb{Z}$ 
with the standard sum and multiplication. On the other hand, the 
simplest example of a finite ring is the set $\mathbb{Z}_n$
of integers modulo $n$, which has exactly $n$ elements.

A field $F$ is a commutative ring with division, that is, such that 0
does not equal 1 and all elements of $F$ except 0 have a multiplicative 
inverse (note that 0 and 1 here stand for the identity elements for the 
addition and multiplication, respectively, which may differ from the 
familiar real numbers 0 and 1). Elements of a field form Abelian 
groups with respect to addition and multiplication (in this latter
case, the zero element is excluded).

The characteristic of a finite field is the smallest integer $p$ such 
that
\begin{equation}
p \, 1= \underbrace{1 + 1 + \ldots + 1}_{\mbox{\scriptsize $p$ times}}=0
\end{equation}
and it is always a prime number. Any finite field contains a prime 
subfield $\mathbb{Z}_p$ and has $d = p^n$ elements, where $n$ is a 
natural number. Moreover, the finite field containing $p^{n}$ elements 
is unique and is called the Galois field $\Gal{p^n}$.

Let us denote as $\mathbb{Z}_p[x]$ the ring of polynomials with
coefficients in $\mathbb{Z}_p$. Let $P(x)$ be an irreducible
polynomial of degree $n$ (i.e., one that cannot be factorized over
$\mathbb{Z}_p$). Then, the quotient space $\mathbb{Z}_p[X]/P(x)$
provides an adequate representation of $\Gal{p^n}$. Its elements can 
be written as polynomials that are defined modulo the irreducible 
polynomial $P(x)$. The multiplicative group of $\Gal{p^n}$ is cyclic
and its generator is called a primitive element of the field.

As a simple example of a nonprime field, we consider the 
polynomial $x^{2}+x+1=0$, which is irreducible in $\mathbb{Z}_{2}$.
If $\prim$ is a root of this polynomial, the elements $\{ 0, 1, \prim , 
\prim^{2} = \prim + 1 = \prim^{-1} \} $ form the finite field $\Gal{2^2}$ 
and $\prim$ is a primitive element.

The map $\alpha \mapsto \alpha^{p}$, where $\alpha \in \Gal{p^n}$ is 
a linear automorphism of $\Gal{p^n}$: $( \alpha + \beta)^{n} =
\alpha^{n} + \beta^{n},$ and $ ( \alpha \beta )^{n} = \alpha^{n}
\beta^{n}$. It is called the Frobenius automorphism and will be
represented in the form
\begin{equation}
 \label{Frob}
 \mathfrak{S}^{k} (\alpha ) = \alpha^{p^k} .
\end{equation}
The elements of the prime field are invariant under action of the this
automorphism.

Another basic map is the trace
\begin{equation}
 \label{deftr}
 \tr (\alpha ) = \alpha + \alpha^{2} + \ldots +
 \alpha^{p^{n-1}} = \sum_{k=0}^{n-1} \mathfrak{S}^{k} ( \alpha ),
\end{equation}
which satisfies
\begin{equation}
 \label{tracesum}
 \tr ( \alpha + \beta ) =
 \tr ( \alpha ) + \tr ( \beta ) ,
\end{equation}
and also leaves the prime field invariant. In terms of it we define 
the additive characters as
\begin{equation}
 \label{Eq: addchardef}
 \chi (\alpha ) = \exp \left [ \frac{2 \pi i}{p}
  \tr ( \alpha ) \right] ,
\end{equation}
and posses two important properties:
\begin{equation}
 \chi (\alpha + \beta ) =
 \chi (\alpha ) \chi ( \beta ) ,
 \qquad \qquad
 \sum_{\alpha \in
  \Gal{p^n}} \chi ( \alpha \beta ) = p^n
 \delta_{0,\beta} .
 \label{eq:addcharprop}
\end{equation}

Any finite field $\Gal{p^n}$ can be also considered as an
$n$-dimensional linear vector space. Given a basis $\{ \basis_{k} \}$,
($k = 1,\ldots, n$) in this vector space, any field element can be
represented as
\begin{equation}
  \label{mapnum}
  \alpha = \sum_{k=1}^{n} a_{k} \, \basis_{k} ,
\end{equation}
with $a_{k}\in \mathbb{Z}_{p}$. In this way, we map each element of
$\Gal{p^n}$ onto an ordered set of natural numbers $\alpha
\Leftrightarrow (a_{1}, \ldots ,a_{n})$.

Two bases $\{ \basis_{1}, \ldots, \basis_{n} \} $ and $\{
\basis_{1}^\prime, \ldots , \basis_{n}^\prime \} $ are dual when
\begin{equation}
  \tr ( \basis_{k} \basis_{l}^\prime ) =\delta_{k,l}.
\end{equation}
A basis that is dual to itself is called selfdual.

There are several natural bases in $\Gal{p^n}$. One is the polynomial
basis, defined as
\begin{equation}
 \label{polynomial}
 \{1, \prim, \prim^{2}, \ldots, \prim^{n-1} \} ,
\end{equation}
where $\prim $ is a primitive element. An alternative is the normal
basis, constituted of
\begin{equation}
 \label{normal}
 \{\prim, \prim^{p}, \ldots, \prim^{p^{n-1}} \}.
\end{equation}
The choice of the appropriate basis depends on the specific problem 
at hand. For example, in $\Gal{2^2}$ the elements $\{ \prim , \prim^{2}\}$ 
are both roots of the irreducible polynomial. The polynomial basis 
is $\{ 1, \prim \} $ and its dual is $\{ \prim^{2}, 1 \}$, while 
the normal basis $\{ \prim , \prim^{2} \} $ is selfdual.


\end{document}